\documentclass[useAMS,usenatbib]{mn2e}
\usepackage{graphicx,amsmath}

\title[A force proportional to velocity squared derived from spacetime algebra]{A force proportional to velocity squared derived from spacetime algebra}
\author[Steen H. Hansen]{Steen H. Hansen,\\
Dark Cosmology Centre, Niels Bohr Institute, University of Copenhagen,
  Jagtvej 128, 2100 Copenhagen, Denmark}
\begin{document}

\pagerange{\pageref{firstpage}--\pageref{lastpage}} \pubyear{2012}

\maketitle

\label{firstpage}

\begin{abstract}
The underlying geometri of spacetime algebra allows one to derive a
force by contracting the relativistic generalization of angular
momentum, ${\bf M}$, with the mass-current, $mw$, where $w$ is a
proper 4-vector velocity.  By applying this force to a cosmological
object, a repulsive inverse distance-square law is found, which is
proportional to the velocity dispersion squared of that structure.  It
is speculated if this finding may be relevant to the recent
suggestion, that such a force may accelerate the expanding universe
with no need for a cosmological constant.
\end{abstract}

%\pacs{95.30.Sf, 96.12.F, 03.30.+, 04.80.C, 95.36.+}


\section{Introduction}

At high velocities space and time are merged together into Minkowski
spacetime, and even though both distances and durations depend on the
observers frame, then the 4-vector $(ct, \vec x)$ has invariant length
under Lorentz transformations.  Similarly the energy-momentum
4-vector, $(\varepsilon/c , \vec p)$, is a proper 4-vector in
Minkowski space. The classical conservation laws, like energy and
momentum conservation arising from symmetries in time and space, thus
have related conservation laws in relativistic physics.

However, for other objects, such as the (polar vector) dynamic mass
moment, $\vec N =  ct \vec p -\varepsilon \vec x/c $, or the (axial
vector) angular momentum, $\vec L = \vec x \times \vec p$, the
corresponding relativistic conservation laws are often not discussed
in detail. One reason for this is that there is no way of combining
$\vec N$ and $\vec L$ into a proper 4-vector.  Instead, one must
combine $\vec N$ and $\vec L$ into an anti-symmetri rank-2 tensor,
$M^{\mu \nu}$ \citep{landau}.

A very similar issue is well-known from electromagnetisme, where the
(polar vector) electric field, $\vec E$, and the (axial vector)
magnetic field, $\vec B$, also combine into an antisymmetri rank-2
tensor, $F^{\mu \nu}$.  When learning about electromagnetism this is
frustrating to some, since most of our intuition is based on the
fields $(\vec E, \vec B)$, however, when performing a Lorentz
transformation, one most often finds oneself performing the
calculations with the physically somewhat less transparant tensor
$F^{\mu \nu}$.

While contemplating the physical meaning of the mathematical space
where $F^{\mu \nu}$ and $M^{\mu \nu}$ live, one is naturally drawn to
spacetime algebra (STA) \citep{hestenes1966, 2003AmJPh..71..691H},
which provides a geometric explanation for the connection between
tensors like $F^{\mu \nu}$ and $M^{\mu \nu}$ and Minkowski space.

In the sufficiently mature scientific field of STA, it is well-known
how the electromagnetic bi-vector field, ${\bf F}$, naturally is used
to derive both Maxwell's equations and the Lorentz force. Given the
strong mathematical similarities between the ``electromagnetic''
bi-vector ${\bf F}$ and the ``mechanics'' bi-vector
${\bf M}$, it appears natural to derive the equations and forces
which are dictated by ${\bf M}$, in particular since the
structure of STA uniquely defines these equations and
forces.

Below the same methods are applied to the bi-vector field, ${\bf M}$,
as have previously been applied to ${\bf F}$ in electromagnetism, and
it is shown how new forces appear naturally from the bi-vector field
${\bf M}$.  One of these inverse distance-squared force-terms depends on the
internal velocity dispersion of an object (which for instance could be
a distant galaxy). It is speculated to which degree this new force possibly may
be related to the force which was recently suggested as an explanation
for the observed acceleration of the
universe~\citep{2021ApJ...910...98L}.

\section{Spacetime algebra}

Spacetime algebra starts with Minkowski space, ${\cal M}_{1,3}$, with
the metric signature $(+,-,-,-)$, and a chosen basis $\{\gamma_\mu
\}_{\mu=0}^3$ of ${\cal M}_{1,3}$. These 4 orthonormal vectors are the
basis for 1-blades. The 2-blade elements are the 6 antisymmetric
products $\gamma _{\mu \nu} \equiv \gamma_{\mu} \gamma_{\nu} $. The
product is here given by the sum of the dot and wedge product: $a b =
a \cdot b + a \wedge b$ \citep{hestenes2015,doranlasenby}.
The wedge operator, $\wedge$, is the 4-dimensional generalization
of the 3-dimensional cross-product.
Continuing over 3-blades, $\gamma _{\mu
  \nu \delta}$, one finally reaches the highest grade, the
pseudoscalar $I \equiv \gamma _0 \gamma _1 \gamma _2 \gamma _3$, which
represents the unit 4-volume in any basis. Interestingly one has $I^2
= -1$.

For the discussion below, the bi-vectors are  important:
these are oriented plane segments, and examples include the
electromagnetic field $ {\bf F} = \vec E + \vec B \, I$, and the
angular momentum ${\bf M} = x \wedge p$, where $x$ and $p$ are proper
4-vectors~\citep{hestenes2015,doranlasenby}.
From a notational point of view vector-arrows are used above spatial 3-vectors
like $\vec E$ or $\vec p$, no-vector-arrows are used for proper 4-vectors like
$x$ and $w$, and boldface is used for bi-vectors like ${\bf F}$ and
${\bf M}$.

\section{Electromagnetism}

The case of electromagnetism in STA is  well described in the
literature \citep{hestenes2015, dressel2015}, and serves as a starting
point here.  The 4 Maxwells equations can be written
\begin{equation}
  \nabla {\bf F} = j \,
  \label{eq:maxwell}
\end{equation}
where the complex current may contain both eletric (vector) and
magnetic (trivector) parts, $j_e + j_m I$. The bi-vector is given by
${\bf F} = \vec E + \vec B I$, and the derivative $\nabla {\bf F} = \nabla
\cdot {\bf F} + \nabla \wedge {\bf F}$ produces both a vector and a
trivector field.

Using the time-direction, $\gamma_0$, one can decompose the derivative along
a direction parallel to and perpendicular to $\gamma_0$,
$\nabla = \left( \partial _0 - \vec \nabla \right) \gamma_0 $,
%\begin{equation}
%\end{equation}
where $ \vec \nabla $ is the frame-dependent relative 3-vector derivative.
It is now straight forward to expand eq.~(\ref{eq:maxwell}) to the
4 Maxwells equation~\citep{hestenes2015, dressel2015}.

It is important to stress, that eq.~(\ref{eq:maxwell}) is not only a
matter of compact notation, it is indeed the only logical extension
beyond the most trivial equation in STA, $\nabla {\bf F} = 0$. The
only thing missing is to connect the bi-vector field to observables:
this is done through observations, which also establish the units of
$\vec E$ and $\vec B$.

\subsection{The Lorentz force}
The classical Lorentz force is given by
\begin{equation}
  \frac{d\vec p}{d t} = q \left(  c \vec E + \vec v \times \vec B \right) \, ,
  \label{eq:lorentz}
\end{equation}
which effectively arose as a clever guess to explain observations.
The cross, $\times$, refers to the normal 3-dimensional cross-product.
In
the standard Euler-Lagrange formalism the Lorentz force appears when
adding a term to the Lagrangian, $q w_\mu A^\mu$, where $A$ is the
4-vector potential.

In STA the Lorentz force appears when one contract the bi-vector field
${\bf F}$ with a proper 4-vector velocity $w$. Using that the 4-vector
$w$ is connected with the para-vector, $w_0+\vec w$, via
$\gamma_0$~\citep{doranlasenby}, namely~\footnote{The
right-multipliation by the timelike vector $\gamma_0$ isolates the
relative quantities of that frame~\citep{dressel2015}, e.g. $x
\gamma_0 = \left( ct + \vec x \right)$.}  $w = \left( w_0 + \vec w
\right) \gamma_0$, one gets
\begin{equation}
  \left(  {\bf F} \cdot w \right) q \frac{d\tau}{dt} \gamma_ 0=
  q  \vec E \cdot \vec v +
  q \left( c \vec E + \vec v \times \vec B\right) \, ,
\label{eq:lorentzforce}
\end{equation}
where the first term on the r.h.s. is the rate of work,
$d\varepsilon/d(ct)$, we use $\vec w = \gamma \vec v$,
and the last parenthesis on the r.h.s is exactly
the Lorentz force in eq.~(\ref{eq:lorentz}).

If the current was complex
%(which could be cause by magnetic monopoles)
there could be another force term allowed~\citep{dressel2015}, namely
\begin{equation}
  {\bf F} \cdot \left( w I \right) \gamma_0 = \left( \left( {\bf F} \wedge w \right) I \right) \gamma_0
    = \vec B \cdot \vec v + \left( c \vec B - \vec v \times \vec E \right) \, .
\end{equation}

To summarize, the full 4 Maxwells equations appear naturally from the
geometric structure of STA, through the equation $\nabla {\bf F} =
j$. The Lorentz force also appears naturally in STA when contracting
the bi-vector field, ${\bf F}$ with the 4-vector current, $dp/d\tau =
{\bf F} \cdot (qw)$, where $p$ is the proper energy-momentum 4-vector.

\section{Angular momentum}
In order to generalize the 3-dimensional angular momentum, $\vec L =
\vec x \times \vec p$, one uses the proper 4-dimensional $x=(ct +\vec
x)\gamma_0$ and $p= (\varepsilon/c + \vec p) \gamma_0$, to create the
bi-vector ${\bf M}$~\citep{landau, dressel2015}
\begin{eqnarray}
  {\bf M} &=& x \wedge p  \nonumber \\
  &=& \frac{\varepsilon \vec x}{c} - ct \vec p - \vec x \times \vec p I  \nonumber \\
  &=& - \vec N - \vec L I  \, .
\end{eqnarray}
Only ${\bf M}$ is a proper geometric object, and the split into
dynamic mass moment and angular momentum requires that one specifies
$\gamma_0$, in exactly the same way that ${\bf F}$ is the proper
geometric object of electromagnetism, and the separation into $\vec E$
and $\vec B$ fields requires specification of a frame by the choice of
$\gamma_0$.

It is now clear how everything can be repeated from the case of
electromagnetism: where one had a bi-vector ${\bf F}$ and relative
3-vectors $\vec E$ (polar) and $\vec B$ (axial), then one now has a
bi-vector ${\bf M}$ and relative 3-vectors $- \vec N$ (polar) and
$-\vec L$ (axial). The signs could have been defined away, but are kept
to agree with the standard notation in the literature~\citep{landau}.  When
deriving the Lorentz force for electromagnetism, by dotting the
bi-vector field ${\bf F}$ with a charge-current, $q w$, one needs
experimental data to get the units right ($\epsilon_0$ and $\mu_0$ for
the E- and B-fields, respectively)~\citep{hestenes2015}. In a similar
fashion experimental data is needed to get the units for a force
defined by dotting the field ${\bf M}$ with a ``mass-current'', $m_t w$.

The simplest possible equation describing the evolution of the
bi-vector field is specified by the structure of STA, namely
\begin{equation}
\nabla {\bf M} = j_{\bf M} \, .
\end{equation}
This Letter is not focusing on the details of the source on the r.h.s. (which could be zero),
however, for the sake of generality it is allowed to contain both a
vector and a trivector term $ j_{\bf M} = j_{\bf 1} + j_{\bf 3} I$.
The resulting equations split into two equations for the relative
scalars
\begin{eqnarray}
-\vec \nabla \cdot \vec N &=& \rho_1  \, , \\
\label{eq:relativescalars} 
- \vec \nabla \cdot \vec L &=& \rho_3 \, ,
\label{eq:relativescalars2}
\end{eqnarray}
(where $\rho_i$ refer to the 0-component of the sources) and two
equations for the relative 3-vectors, just like Maxwell's equations
did.
\begin{eqnarray}
  -\partial_0 \vec N + \vec \nabla \times \vec L &=& - \vec J_1 \, ,\\
  \partial_0 \vec L + \vec \nabla \times \vec N &=&  \vec J_3  \, ,
  \label{eq:relvec2}
\end{eqnarray}
where $\vec J_i$ refer to the 3 spatial components of the sources.
The details of these 4 equations
will be discussed elsewhere~\citep{students}.

Instead, the force which appears from the
contraction with a mass-current, $m_t \omega$, where $w$ again is a
proper 4-velocity, and $m_t$ is the inertia of the test particle,
will now be calculated.  From the term ${\bf M} \cdot w$ one gets
\begin{equation}
  \left( {\bf M} \cdot w  \right) \frac{d\tau}{dt} \gamma_0 = -\vec N \cdot \frac{\vec v}{c} +
  \left( -  \vec N - \frac{\vec v}{c} \times \vec L \right) \,.
  \label{eq:newforce}
\end{equation}
The first term on the r.h.s. is similar to a rate of work. However,
the last parenthesis of eq.~(\ref{eq:newforce})
contains the new forces of interest here,
and
will be discussed in section \ref{sec:newforce}
below.  One could also have considered a force arising from $\left(
{\bf M} \wedge w \right) I$, which looks like
\begin{equation}
\left( {\bf M} \wedge w \right) I \frac{d\tau}{dt} \gamma_0 =
   - \vec L \cdot \frac{\vec v}{c}
  + \left( -\vec L + \frac{\vec v}{c} \times \vec N \right) \, ,
  \label{eq:newforce2}
\end{equation}
however, it is left for a future analysis to study this.
%in any detail~\citep{students}.

\section{The new force terms}
\label{sec:newforce}

Let us consider a collection of particles at a large distance, $\vec
r_0$.  The particles may have different inertia, $m_i$, but move
collectively with an average velocity, $\vec V$. If one considers
particles in a cosmological setting, then the velocity is a
combination of the Hubble expansion and peculiar velocity, $\vec V = H
\vec r_0 + \vec v_p$, and at large distances the peculiar velocity is
subdominant. The collection of particles may have internal motion,
which is simplified with an internal velocity dispersion,
$\sigma^2$. Practically when calculating the velocity
dispersion there will be terms including both the Hubble expansion,
$v_H = H r$, and also the background density of both matter and the
cosmological constant, however, these terms happen to exactly cancel
each other~\citep{2013MNRAS.431L...6F}, and one can therefore
calculate $\sigma^2$ as if the structure is alone in a non-expanding
universe.

The dynamic mass moment is given by
\begin{equation}
\vec N = \sum \left( ct \vec p_i - \frac{\varepsilon_i \vec r_i}{c} \right) \, ,
\end{equation}
where the sum is over all particles involved~\citep{landau}. If one
divides both terms by the total energy, $\varepsilon_{tot} = \sum
\varepsilon_i$, then one gets
\begin{equation}
  \frac{\vec N }{\varepsilon_{tot}} = \left( \frac{ct \sum \vec
    p_i}{\sum \varepsilon_i} - \frac{\sum \varepsilon_i \vec r_i}{c \sum
    \varepsilon_i} \right) \, ,
\end{equation}
The first term is just $ct$ times the average velocity. At small velocities
one has $\varepsilon \approx m_ic^2$ and hence the last term describes
the relativistic center of inertia, $\vec R_{cm} = \sum (m_i \vec
r_i)/ \sum m_i$.
If the centre of inerti moves at constant velocity (now ignoring sums over
particles), then one has $\vec r = \vec r_0 + \vec V t$, and hence
\begin{equation}
  \vec N = - m c \vec r_0 \,  .
\label{eq:Nr}
\end{equation}

When considering the Lorentz force in eq.~(\ref{eq:lorentzforce}) one needs
to get the units right to get 
$\vec E = q \vec r /(4 \pi \epsilon_0 |r|^3)$, which includes
the observable vacuum permittivity, $\epsilon_0 $, and also Coulombs
inverse distance-square law (resulting from Gauss' and Faradays' laws combined).
Effectively this means dividing by $\epsilon_0 |r|^2$.

The new force terms in the parenthesis of eq.~(\ref{eq:newforce}) will
now be considered.  Since the mass-current is related to gravity, one
should multiply by Newtons gravitational constant, $G = 6.67 \times
10^{-11} {\rm m}^3/({\rm s}^2 \, {\rm kg})$.  To get a well-behaved
field one divides by distance to power 3, which will lead to an inverse
distance-square force: this comes from the integral over
eq.~(\ref{eq:relativescalars}), using the sphericity from
eq.~(\ref{eq:relvec2}) with $\vec J_3=0$ and the expression in
eq.~(\ref{eq:Nr}). This last point is easily recognized by considering
the change of notation, $\vec N \rightarrow \vec g /(4 \pi G)$, and
$\rho_1 \rightarrow \rho_m$ which is the mass density, which means
that eq.~(\ref{eq:relativescalars}) is written as $\vec \nabla \cdot
\vec g = - 4 \pi G \rho_m$. This equation is clearly regognized as
leading to Newtons gravitational law.  Finally to get the units right,
it is divided by $c$.

This implies that one has a force-term that looks like
\begin{equation}
  - \kappa \, \frac{G m_t m \vec r_0} {| \vec r_0 |^3} \, .
\label{eq:newton}
\end{equation}
where $\kappa$ is
an unknown, dimensionless number, which must be
determined from observations,
and $m_t$ is from the
mass-current, $m_t w$.
In the case of $\kappa = 1$ this is
just Newtons gravitational force.
%The classical approach to get the Newtonian gravity is to introduce a
%potential in the Lagrangian , ${\cal L} = 1/2 \, m v^2 - m \phi$, in
%such a way that the force appears as the derivative of the potential,
%$\dot{ \vec v }= - \vec \nabla \phi$. Here, i
In the above picture, it thus appears that the Newtonian gravity may
be interpreted as a gravitational analogue to the Coulomb force from
electromagnetism. Since the masses are always positive, the
gravitational force is always attractive.

When the structure under consideration contains a dynamical term
proportional to the velocity dispersion, $\sigma^2$, which for
instance can arise in a dwarf galaxy where the stars and dark matter
particles are orbiting in the local gravitational potential, then the
potential will be minus 2 times the kinetic energy according to the
virial theorem~\citep{bt2}, $2T+U=0$, and hence one writes the
energy as
\begin{equation}
\varepsilon_i = m_ic^2 - \frac{1}{2} m_i \sigma_i^2 \, .
\end{equation}
In this case one ends up with a new force of the form 
\begin{equation}
   \frac{\tilde \kappa}{2} \, \frac{\sigma^2}{c^2} \,
    \frac{m_tm_iG \vec r_0}{\| \vec r_0 \|^3} \, ,
\label{eq:sigma2}
\end{equation}
where the dispersion has been  normalized to $c$.  This force is always
repulsive. In the case of $\tilde \kappa =1$ this is just a minor
correction to the normal gravitational attraction, e.g. for galaxy
clusters with velocity dispersions of $1000$ km/sec this is a
$10^{-5}$ correction, and for dwarf galaxies much less. Possibly for
motion near very compact objects, this correction may eventually be
observable.

Velocity-dependent forces are well-known, including the Coriolis-force
and the Lorentz-force, however, this is, to our knowledge, the first
derived long-distance force depending on velocity squared.  In an
atttempt to make kinetic energy depend on relative velocities (rather
than absolute velocities) similarly to how potential energy depends on
relative position, Schr\"odinger suggested a new gravitational force
proportional to velocity squared \citep{1925AnP...382..325S}. His
force has essentially the same form as the force derived above,
however, its existence was postulated on rather philosofical grounds.

A recent paper demonstrated that a universe which contains no dark
energy, but instead includes a new repulsive inverse  distance-square force
proportional to internal velocity dispersion squared, just like
equation~(\ref{eq:sigma2}), could have an accelerated expansion which
fairly closely mimics the accelerated expansion induced by the
cosmological constant~\citep{2021ApJ...910...98L}.  In that paper it
was implicitely suggested that such a force might conceivably exist amongst the
dark matter particles. What has been shown in this Letter is, that
such a force indeed may exist, and that it is not specifically related
to the dark matter particle, but instead related to gravity in general.
One of the concerns with the suggestion discussed in
\cite{2021ApJ...910...98L} is the potential instability of
cosmological structures, however, from the derivation above it is clear
that the new force derived here
comes from the internal dispersion (as opposed to relative velocities),
and hence there is no instability concern.

The magnitude of the dimensionless $\tilde \kappa$ in
eq.~(\ref{eq:sigma2}) is unknown in the present derivation, however,
it should logically be unity.  The new force of the
paper~\citep{2021ApJ...910...98L} should have a numerical value of
$\tilde \kappa \sim 10^6-10^8$. From the derivation above there is no
indication where such a large factor should come from.

\section{Conclusion}
It was recently suggested that if a force which depends on velocity
squared exists in Nature, then it may induce an effect on cosmological
scales which mimics the accelerated expansion of the standard
cosmological constant~\citep{2021ApJ...910...98L}. The present Letter
demonstrates one such concrete possibility.  The derivation here is
framed in the geo\-metric structure of spacetime
algebra~\citep{hestenes2015}, and takes as starting point the
relativistic generalization of angular momentum which includes the
dynamic mass moment, $\vec N = ct \vec p - \varepsilon \vec
r/c$. Since the energy in this term, $\varepsilon$, contains the
velocity dispersion of a distant cosmological object, then an inverse
distance-square force naturally appears, which is proportional to
$\sigma^2$.
Such a force 
may lead to a slightly reduced
gravitational force for extremely compact objects.
The magnitude of the force derived here is significantly
smaller than needed to explain the present day accelerating
universe~\citep{2021ApJ...910...98L}.

\section*{Acknowledgement}
It is a pleasure to thank Max Emil K.S. Sondergaard, Magnus B. Lyngby
and Nicolai Asgreen for interesting discussions.  I thank Mario
Pasquato for bringing the 1925 Schr\"odinger paper to my attention.

\section{Data availability}
No new data were generated or analysed in support of this research.

\label{lastpage}

\end{document}